\documentclass[sn-nature]{sn-jnl}

\usepackage{graphicx}%
\usepackage{multirow}%
\usepackage{amsmath,amssymb,amsfonts}%
\usepackage{amsthm}%
\usepackage{mathrsfs}%
\usepackage[title]{appendix}%
\usepackage{xcolor}%
\usepackage{textcomp}%
\usepackage{manyfoot}%
\usepackage{booktabs}%
\usepackage{algorithm}%
\usepackage{algorithmicx}%
\usepackage{algpseudocode}%
\usepackage{listings}%

\usepackage{lineno}
\usepackage{siunitx}
\usepackage{geometry}
\usepackage[normalem]{ulem}
\theoremstyle{thmstyleone}%
\theoremstyle{thmstyletwo}%

\theoremstyle{thmstylethree}%

\begin{document}

\title[Article Title]{Laboratory realization of relativistic pair-plasma beams}


\author*[1]{\fnm{C.~D.} \sur{Arrowsmith}}\email{charles.arrowsmith@physics.ox.ac.uk}
\author[2,3]{\fnm{P.} \sur{Simon}}
\author[4]{\fnm{P.} \sur{Bilbao}}
\author[1]{\fnm{A.~F.~A.} \sur{Bott}}
\author[2]{\fnm{S.} \sur{Burger}}
\author[5]{\fnm{H.} \sur{Chen}}
\author[4]{\fnm{F.~D.} \sur{Cruz}}
\author[6]{\fnm{T.} \sur{Davenne}}
\author[2]{\fnm{I.} \sur{Efthymiopoulos}}
\author[7]{\fnm{D.~H.} \sur{Froula}}
\author[2]{\fnm{A.~M.} \sur{Goillot}}
\author[8,9]{\fnm{J.~T.} \sur{Gudmundsson}}
\author[7]{\fnm{D.} \sur{Haberberger}}
\author[1]{\fnm{J.} \sur{Halliday}}
\author[1,10]{\fnm{T.} \sur{Hodge}}
\author[1]{\fnm{B.~T.} \sur{Huffman}}
\author[1]{\fnm{S.} \sur{Iaquinta}}
\author[1]{\fnm{F.} \sur{Miniati}}
\author[11]{\fnm{B.} \sur{Reville}}
\author[1]{\fnm{S.} \sur{Sarkar}}
\author[1]{\fnm{A.~A.} \sur{Schekochihin}}
\author[4]{\fnm{L.~O.} \sur{Silva}}
\author[5]{\fnm{R.} \sur{Simpson}}
\author[1,2,12]{\fnm{V.} \sur{Stergiou}}
\author[6]{\fnm{R.~M.~G.~M.} \sur{Trines}}
\author[11]{\fnm{T.} \sur{Vieu}}
\author[2]{\fnm{N.} \sur{Charitonidis}}
\author[6,13]{\fnm{R.} \sur{Bingham}}
\author[1]{\fnm{G.} \sur{Gregori}}

\affil[1]{\orgdiv{Department of Physics}, \orgname{University of Oxford}, \orgaddress{\street{Parks Road}, \city{Oxford}, \postcode{OX1 3PU}, \country{UK}}}

\affil[2]{\orgname{European Organization for Nuclear Research (CERN)}, \orgaddress{\street{CH-1211 Geneva 23}, \country{Switzerland}}}

\affil[3]{\orgname{GSI Helmholtzzentrum für Schwerionenforschung GmbH}, \orgaddress{\street{Planckstraße 1
64291 Darmstadt}, \country{Germany}}}

\affil[4]{\orgdiv{GoLP/Instituto de Plasmas e Fusão Nuclear, Instituto Superior Técnico}, \orgname{Universidade de Lisboa}, \orgaddress{\postcode{1049-001}, \city{Lisboa}, \country{Portugal}}}

\affil[5]{\orgname{Lawrence Livermore National Laboratory}, \orgaddress{\street{7000 East Ave}, \city{Livermore}, \state{California}, \postcode{94550}, \country{USA}}}

\affil[6]{\orgname{Rutherford Appleton Laboratory}, \orgaddress{\street{Chilton}, \city{Didcot}, \postcode{OX11 0QX}, \country{UK}}}

\affil[7]{\orgname{University of Rochester Laboratory for Laser Energetics}, \orgaddress{\city{Rochester}, \state{NY},\postcode{14623}, \country{USA}}}

\affil[8]{\orgdiv{Science Institute}, \orgname{University of Iceland}, \orgaddress{\street{Dunhaga 3}, \postcode{IS-107}, \city{Reykjavik},  \country{Iceland}}}

\affil[9]{\orgdiv{School of Electrical Engineering and Computer Science}, \orgname{KTH Royal Institute of Technology}, \orgaddress{ \postcode{SE-100 44}, \city{Stockholm},  \country{Sweden}}}

\affil[10]{\orgname{AWE}, \orgaddress{\street{Aldermaston}, \city{Reading}, \state{Berkshire}, \postcode{RG7  4PR}, \country{UK}}}

\affil[11]{\orgname{Max-Planck-Institut für Kernphysik}, \orgaddress{\street{Saupfercheckweg 1}, \postcode{D-69117}, \city{Heidelberg}, \country{Germany}}}

\affil[12]{\orgdiv{School of Applied Mathematics and Physical Sciences}, \orgname{National Technical University of Athens}, \orgaddress{\postcode{Athens 157 72}, \country{Greece}}}

\affil[13]{\orgdiv{Department of Physics}, \orgname{University of Strathclyde}, \orgaddress{\city{Glasgow}, \postcode{G4 0NG}, \country{UK}}}

\maketitle

\centerline{\large(Dated: \today)}

\newpage

{\bf Relativistic electron-positron (e$^{\pm}$) plasmas are ubiquitous in extreme astrophysical environments such as black holes and neutron star magnetospheres, where accretion-powered jets and pulsar winds are expected to be enriched with such pair plasmas. Their behaviour is quite different from typical electron-ion plasmas due to the matter-antimatter symmetry of the charged components and their role in the dynamics of such compact objects is believed to be fundamental. So far, our experimental inability to produce large yields of positrons in quasi-neutral beams has restricted the understanding of electron-positron pair plasmas to simple numerical and analytical studies which are rather limited. We present first experimental results confirming the generation of high-density, quasi-neutral, relativistic electron-positron pair beams using the 440 GeV/c beam at CERN's Super Proton Synchrotron (SPS) accelerator. The produced pair beams have a volume that fills multiple Debye spheres and are thus able to sustain collective plasma oscillations. Our work opens up the possibility of directly probing the microphysics of pair plasmas beyond quasi-linear evolution into regimes that are challenging to simulate or measure via astronomical observations.}


\section*{Main}\label{Introduction}

Relativistic electron-positron (e$^{\pm}$) pair plasmas are expected to be produced around black holes \cite{Bambi:2016lkv}
and neutron stars \cite{arons1979some}. 
In these environments, pair creation can occur due to intense, high-energy $\gamma$-ray fluxes (by the Breit-Wheeler process \cite{breit1934collision}) or when the electromagnetic fields are comparable to the Schwinger field: the critical field strength for vacuum breakdown ($E_{\rm c} = \SI{1.3e18}{\volt}$/m, $B_{\rm c}=\SI{4.4e9}{\tesla}$) \cite{schwinger1951gauge,erber1966high}.
Because of the symmetry of the charged components, electron-positron pair plasmas should exhibit collective behaviour that is significantly different from typical electron-ion plasmas~\cite{tsytovich1978laboratory}. Linear and non-linear wave processes can be affected in both fluid and kinetic regimes because of the suppression of some wave modes. This is important in a variety of astrophysical settings, with recent attention focusing on fast radio burst generation and the stability of astrophysical pair beam jets~\cite{arons1983pair,begelman1984theory,blandford1995pair,turolla2015magnetars,lyubarsky2021emission}.
However, producing sufficiently large yields and densities of e$^{\pm}$ pairs in the laboratory in order to directly probe the relevant plasma microphysics has been challenging. Presently, high flux laboratory sources of positrons include: (i) nuclear reactors \cite{hugenschmidt2012nepomuc}, (ii) electron accelerators \cite{bernardini2004ada,blumer2022positron}, and (iii) high-power lasers \cite{chen2015scaling,liang2015high,sarri2015generation,xu2016ultrashort,peebles2021magnetically,jiang2021enhancing,chen2023perspectives}. All these approaches involve pair production processes when sufficiently energetic $\gamma$-rays ($E_{\gamma}\geq 2 m_{\rm e} c^2 = \SI{1.022}{\mega\electronvolt}$) interact with charged nuclei (so-called Trident and Bethe-Heitler processes \cite{bethe1934stopping}), with the highest cross-section in high-Z materials. In the coming decade, it is proposed to use magnetic chicanes at FACET-II (SLAC) to combine the accelerator's e$^+$ and e$^-$ beams into a quasi-neutral jet \cite{yakimenko2019facet}. The next generation of ultra-intense lasers may also be able to produce pairs by achieving the Schwinger limit for vacuum breakdown \cite{bell2008possibility,ridgers2012dense,zhang2020relativistic}. Meanwhile, precision magnetic confinement techniques have been developed to trap low-temperature e$^{\pm}$ pair plasmas \cite{greaves1995electron,danielson2015plasma,stenson2018lossless}, and relativistic laser-produced plasmas \cite{chen2014magnetic,von2021confinement,stoneking2020new}. 
However, despite significant efforts, none of these approaches have so far been able to produce the yields and densities of pairs needed to sustain collective modes in the plasma.



\begin{figure*}[t]
\centering
\includegraphics[width=0.99\textwidth]{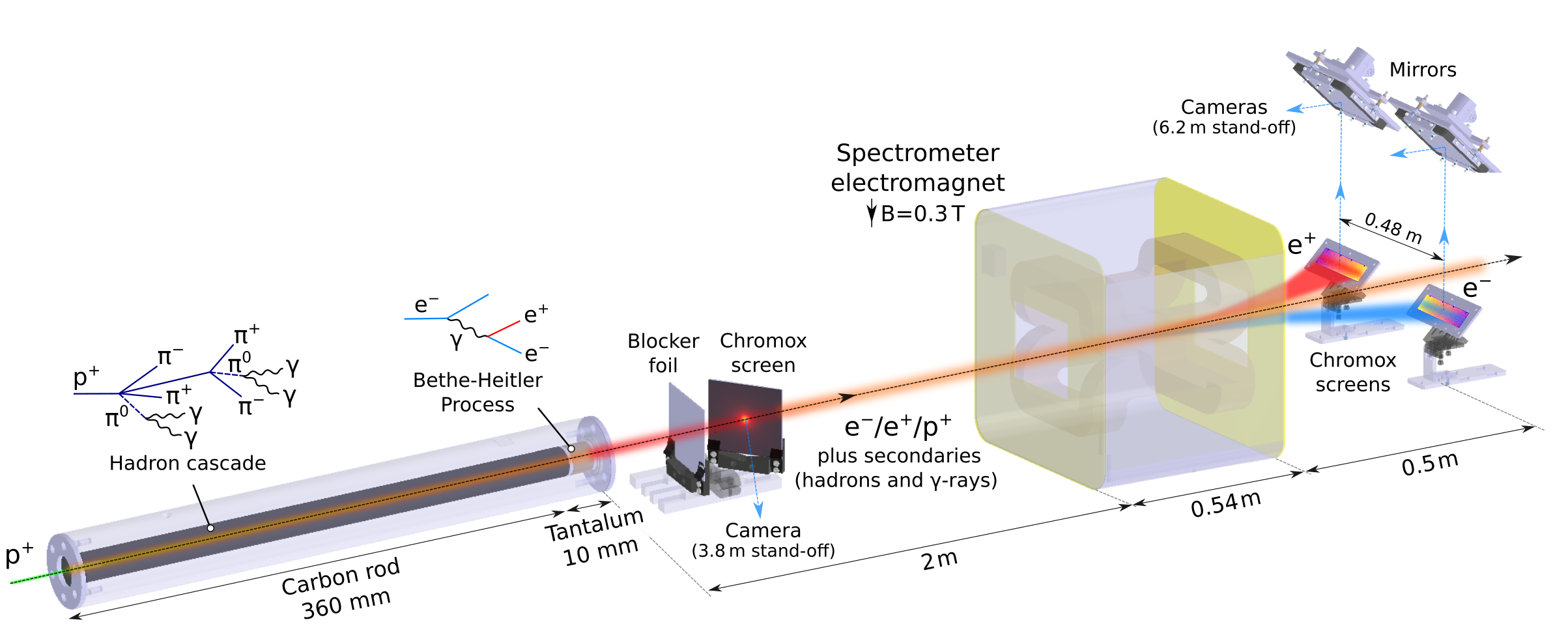}
\caption{{\bf Experimental setup.} 
Protons with 440 GeV/c momentum are extracted from the SPS ring with maximum fluence exceeding $3\times10^{11}$ protons in a single bunch of duration \SI{250}{\pico\second} (1-$\sigma$), and transverse size $\sigma_r = \SI{1}{\milli\metre}$. The transverse beam profile of the secondary beam is imaged using a $70$~mm~$\times\, 50$~mm~$\times \, 0.25$~mm chromium-doped (Chromox) luminescence screen positioned 10~cm downstream of the target, and a blocker foil (\SI{50}{\micro\metre} aluminium) is used to minimize stray optical light. The Chromox plate is oriented at 45$^{\circ}$ to the beam path and viewed by a digital camera which has an exposure time of 24~ms to capture the entire scintillation of the screen.
The \SI{3.8}{\metre} standoff distance of the digital camera leads to image resolution of \SI{50}{\micro\metre}, however the actual resolution is \SI{100}{\micro\metre} due to the translucence of the Chromox.
At a distance \SI{2}{\metre} downstream of the target, electrons and positrons are separated from the secondary beam and spectrally resolved using a magnetic spectrometer comprised of an electromagnet and a pair of luminescence screens ($200$~mm~$\times\, 50$~mm~$\times \, 1$~mm) centred at a distance \SI{240}{\milli\metre} off-axis. 
20-cm thick bricks of concrete (not shown in the diagram) are placed at the entrance of the electromagnet, leaving a \SI{40}{\milli\metre} wide aperture. Concrete is also placed to block the target from the direct view of the cameras to minimize speckle background arising on the camera images from the impact of high energy hadrons scattered around the experimental area.}
\label{fig:Experimental_Setup}
\end{figure*}

Here we present a novel approach for producing quasi-neutral e$^{\pm}$ jets in which a high-intensity, ultra-relativistic proton beam is converted into pairs via hadronic and electromagnetic cascades with 2-3 orders of magnitude higher yield than previously reported neutral beams \cite{sarri2015generation,peebles2021magnetically}. We performed our experiment at the HiRadMat (High-Radiation to Materials) facility \cite{efthymiopoulos2011hiradmat} in the accelerator complex at CERN, Geneva. The experimental setup is shown in Figure~\ref{fig:Experimental_Setup}. We have performed detailed Monte-Carlo simulations using the 
standard computer code FLUKA \cite{ahdida2022new,battistoni2015overview,vlachoudis2009flair} to characterize the e$^{\pm}$ pair production, as well as the other secondary beam components (hadrons and $\gamma$-rays). The predicted number of pairs produced with kinetic energy greater than \SI{1}{\mega\electronvolt} is $N_{\pm} = \frac{1}{2}(N_{\rm{e}^+}+N_{\rm{e}^-})=1.5\times10^{13}$, with peak pair density $n_{\pm} = \SI{1.6e12}{\per\centi\metre\cubed}$, and the ratio of positrons to electrons $N_{\rm{e}^+}/N_{\rm{e}^-}=0.82$. Downstream of the target, the positron ratio is even higher ($N_{\rm{e}^+}/N_{\rm{e}^-}\gtrsim0.9$), as low-energy electrons present in the beam at the target rear surface preferentially escape the beam due to their higher divergence.

The large numbers of electron-positron pairs are generated using a single LHC-type bunch of $3\times10^{11}$ protons with momentum $440~\mbox{GeV/c}$ and duration (1-$\sigma$) of $\tau=\SI{250}{\pico\second}$. The protons are extracted to the facility from the Super Proton Synchrotron, irradiating a solid-target composed of a low-Z material (graphite) and a high-Z converter (tantalum). The dominant process for producing electron-positron pairs is hadronization of quarks and gluons inside the graphite section of the target. This produces a shower of pions, kaons and other hadrons on scales comparable to the nuclear interaction length in graphite \cite{arrowsmith2021generating}. A copious number of ultra-relativistic neutral pions are produced which almost instantaneously undergo electromagnetic decay to produce a highly-collimated flux of GeV-scale $\gamma$-rays. Electromagnetic cascades are then generated with the $\gamma$-rays producing pairs in the high-Z tantalum converter which is much longer than the conversion length. Further pairs are created via subsequent bremsstrahlung of electrons and positrons (Bethe-Heitler process). Secondary $\gamma$-rays which do not convert into pairs can escape the target, along with a much smaller (by orders of magnitude) number of protons and other hadronic species. The effect of hadronic beam components (such as $\pi^{\pm}$ pairs) on e$^{\pm}$ pair plasma dynamics must be considered, but the effects are expected to be negligible due to the much lower mobility and density of these species. 
The choice of target material length constitutes a compromise between the number of pairs produced and the emittance of the e$^{\pm}$ beam, with the thickness of graphite and tantalum chosen in the current setup to maximize the pair density ($n_{\pm}$) maintained over a \SI{1}{\metre} length downstream of the target. A choice of thicker target materials can produce an even greater pair density and number of electron-positron pairs at the immediate rear of the target \cite{arrowsmith2021generating}.

Measurements have been performed to validate the predicted beam characteristics by recording the transverse beam profile and particle energy spectra (see Figure~\ref{fig:Experimental_Setup} for details).
This is achieved using 
chromium-doped alumina-ceramic (Chromox) luminescence screens \cite{mccarthy2002characterization,burger2016scintillation,gorgisyan2018commissioning} placed in the beam path downstream of the target. When ionizing particles or radiation are incident on the screen, it emits red visible light with a few ms decay time. 
For our conditions, the intensity of light emitted from the screen is directly proportional to the energy deposited by ionizing particles (see Methods). Although it is not possible to distinguish which secondary particle species is the cause of observed luminescence light, above relativistic energies (Lorentz factor, $\Gamma\gtrsim 2$) charged particles deposit energy as `minimum-ionizing', a characteristic which results in relativistic charged particles depositing an almost identical amount of energy into the screen (see Supplementary Information). Given that the vast majority of secondary particles incident on the screen are relativistic, the observed brightness is thus proportional to the number of incident particles. The expectation from FLUKA simulations 
is that the production of $10^{13}$ pairs will lead to about 100 times larger luminescence compared with a primary beam containing $\sim10^{11}$ protons. 
Since the target is mounted onto a motorized stage, it can be entirely removed from the proton beam path. By independently measuring the incident proton beam fluence for each shot using upstream current monitors, an absolute calibration of the charge incident on the screen can be made. 

A common source of background in experiments producing energetic electromagnetic cascades is the large number of scattered sub-MeV e$^-$ and $\gamma$-rays which can flood the detectors. Given that our experiment is carried out in an air environment, a large fraction of low-energy particles and radiation are absorbed before they can reach the screens; for instance, e$^\pm$ with energy $\lesssim\SI{100}{\kilo\electronvolt}$ are mostly absorbed by a few centimetres of air. The air environment can provide additional sources of background in the form of stray light arising from Cherenkov emission, fluorescence of air molecules, and optical transition radiation (OTR) which is produced by particles passing from different dielectric materials into the air (such as the target-air interface). While the contribution of all of these sources is small compared to the light collected due to luminescence, an aluminium blocker foil is placed in the beam path before the luminescence screen to reduce on-axis Cherenkov and OTR from illuminating the Chromox plate.  
\\

\begin{figure*}[t]
\centering
\includegraphics[width=1\textwidth]{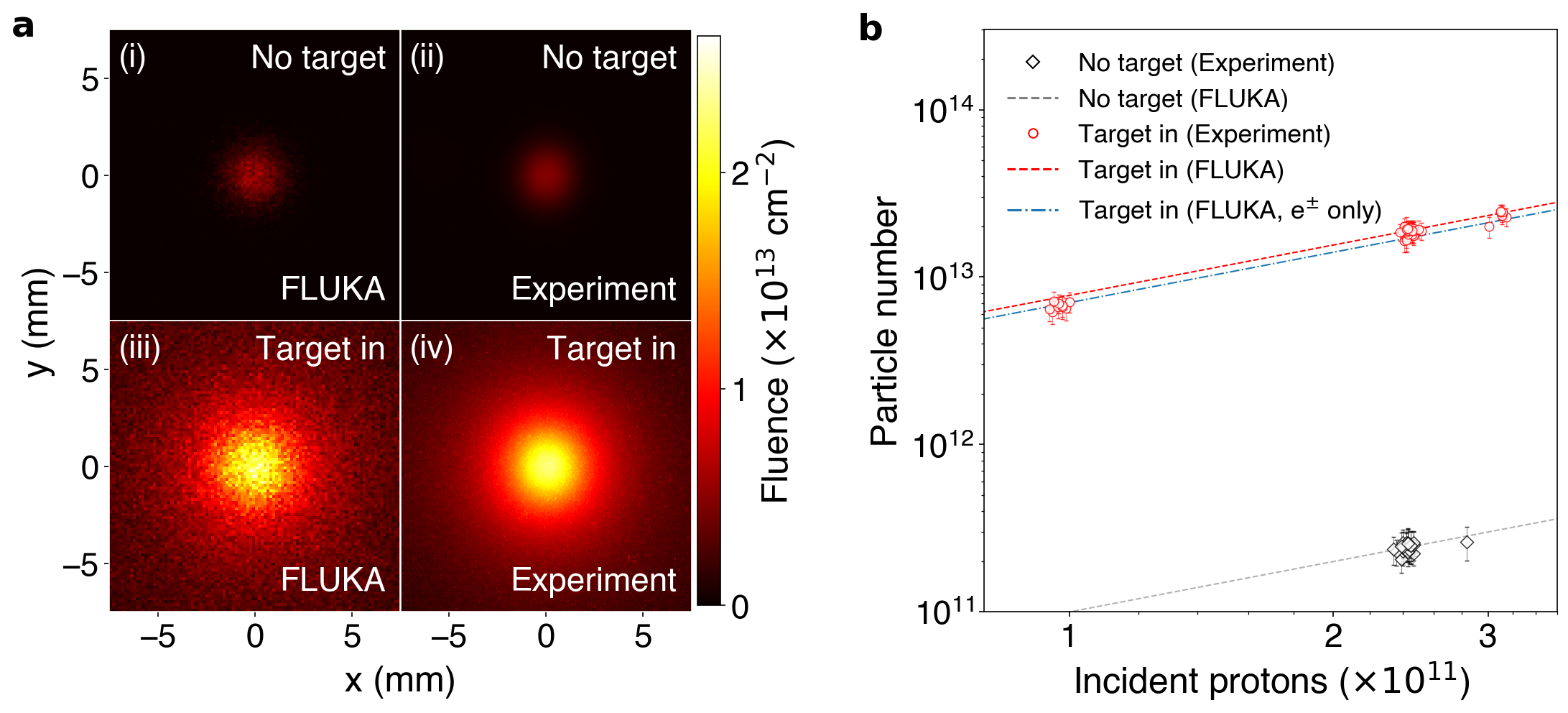}
\caption{{\bf Transverse beam profile imaged using a luminescence screen.} (a) Direct comparison of FLUKA Monte-Carlo simulations with raw image data obtained when the target is irradiated and the secondary beam is produced (`Target in'), versus when the target is removed and only the primary proton beam irradiates the screen (`No target'). An absolute fluence calibration is obtained using the known density profile of the primary proton beam. (b) Integrated image intensity (total intensity) from 70 shots is converted to an absolute particle number, showing the case where the target is irradiated (red circles, 47 shots), and when it is removed (black diamonds, 23 shots). The error bars reflect the standard errors of the fitted parameters for each shot. FLUKA Monte-Carlo simulations of the predicted light yield are shown for both cases (black-dashed and red-dashed lines), showing strong agreement with the experimental data. The blue dot-dashed line shows the contribution from e$^{\pm}$ in the FLUKA simulation, highlighting that this is the dominant contribution to the enhanced signal.}
\label{HRM3_figure_lowres_v4}
\end{figure*}

The experimental results of the post-target in-beam luminescence screen are summarized in Figure~\ref{HRM3_figure_lowres_v4}, comparing directly with FLUKA Monte-Carlo simulations. Figure~\ref{HRM3_figure_lowres_v4}a shows the raw image data of the transverse beam profile when the target is irradiated compared with when the target is removed from the proton path. The image intensity is converted to an absolute particle fluence by normalizing to the known beam density profile of the primary proton beam (intensity $3\times10^{11}$ protons, Gaussian width $\sigma_r = \SI{1}{\milli\metre}$, temporal duration $\tau=\SI{250}{\pico\second}$). When the target is irradiated and the secondary beam is produced, a 5 times increase in peak brightness is observed. In addition, the transverse size of the beam broadens to a Lorentzian profile with half-width $\Sigma_r = 2.3$~mm due to its increased divergence from Coulomb scattering of pairs with atomic nuclei in the target material.

The integrated image intensities (total intensities) obtained from 70 shots with the target irradiated (red circles) and the target removed (black diamonds) are shown in Figure~\ref{HRM3_figure_lowres_v4}b, again converted to an absolute particle number. The total intensity scales with the number of protons in the primary beam. Given that $90$\% of the measured signal corresponds to e$^-$ and e$^+$, the number of e$^{\pm}$ captured on the screen is measured to be $N_{\pm,\rm{exp}}=(1.02\pm0.05)\times10^{13}$. This measurement agrees with the FLUKA expectation for a screen placed the same distance downstream of the target, $N_{\pm,\rm{FLUKA}}=1.04\times10^{13}$. We estimate the peak pair density at the screen position to be $n_{\pm,\rm{exp}} = \SI{9e11}{\per\centi\metre\cubed}$, taking the longitudinal beam profile as identical to the primary beam (beam elongation due to straggling in the target is calculated to be $\lesssim2\%$).
\\

\begin{figure}[t]
\centering
\includegraphics[width=0.5\textwidth]{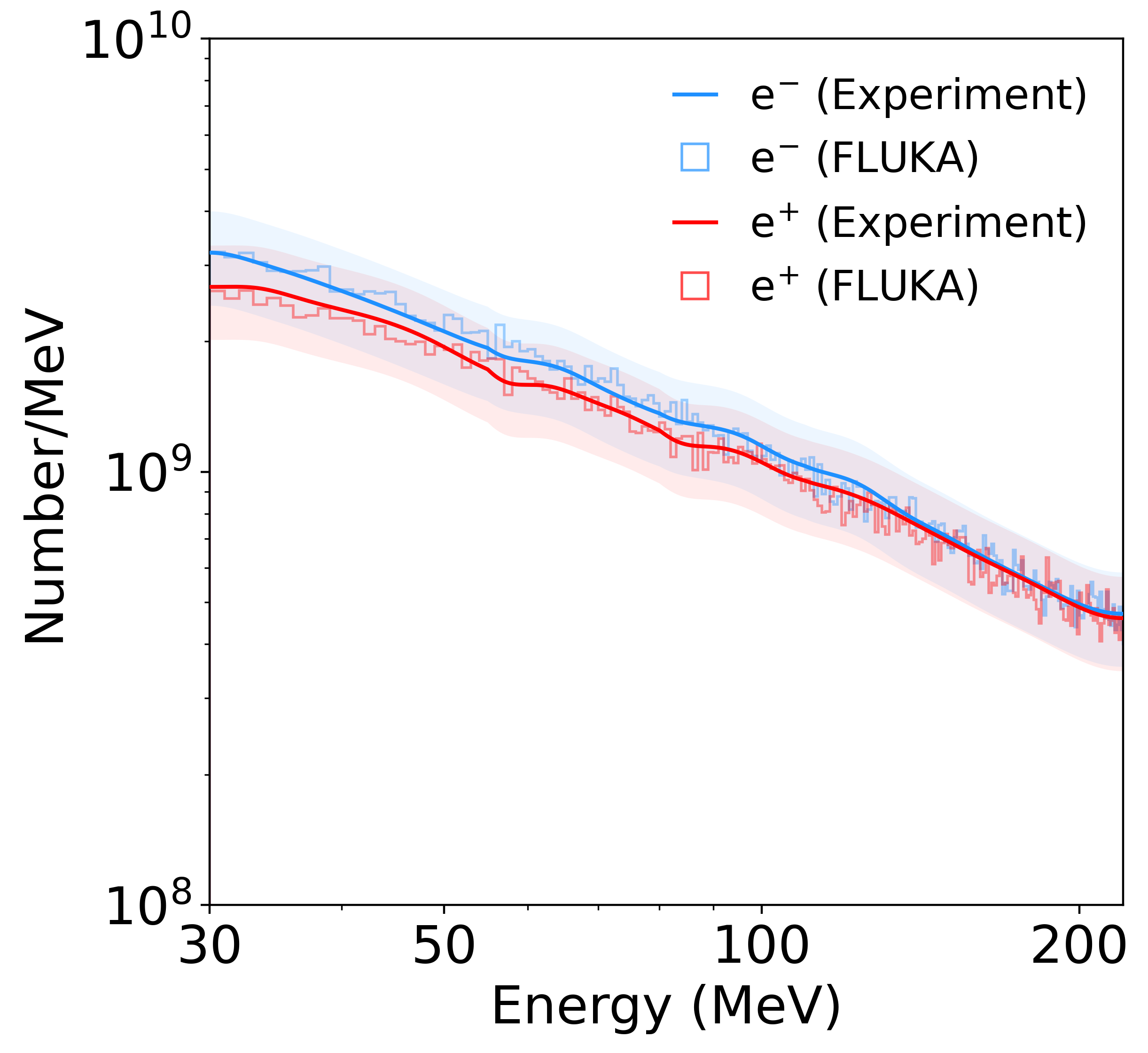}
\caption{{\bf Magnetic particle spectrometer.} The energy spectra of e$^-$ (blue) and e$^+$ (red) are obtained from images of luminescence screens (dimensions $200$~mm~$\times\,50$~mm~$\times\,1$~mm) centred \SI{240}{\milli\metre} from the beam axis on either side (see Figure~1). Electrons and positrons are deflected onto the screens by the vertically-oriented dipole magnetic field of the electromagnet, whilst hadrons with a higher momentum and uncharged $\gamma$-rays are mostly absorbed by the beam dump behind the electromagnet. The spectrum in the energy range $30\leq E~\mbox{[MeV]}\leq220$ is constructed by piecing together images from multiple shots using different magnetic field strengths ($B=0.1-$\SI{0.34}{\tesla}). The shaded regions correspond to the error associated with the absolute calibration. FLUKA simulations (histograms) are able to accurately predict the experimentally obtained spectra.}
\label{fig:HRM5+6}
\end{figure}

An electromagnet is used to measure the particle energy spectra of the electrons and positrons separately from the other secondaries. 
The current in the magnet coils can be varied for different shots to sample different portions of the e$^-$ and e$^+$ energy spectra between $30-220$~MeV with an energy resolution $10-20$\%. In addition, the electromagnet can be switched off, making it possible to characterize the $\gamma$-ray and hadron background. FLUKA simulations predict that without the magnetic field, low energy ($\lesssim10$~MeV) scattered e$^{\pm}$ can irradiate the screens, but when the electromagnet is activated, these lower-momentum pairs are swept away and the measured signals more closely resemble the true spectra. This is confirmed by the overlapping of the segments of the spectrum measured in different energy ranges when the full spectrum is pieced together (shown in Figure~\ref{fig:HRM5+6}). At positions on the luminescence screens sufficiently far from the beam axis ($\gtrsim200$~mm) the number of particles irradiating the screens is enhanced compared with when the electromagnet is deactivated (raw data is shown in the Supplementary Information). This observation can only be explained by the presence of large numbers of e$^{\pm}$ with momentum $\leq220$~MeV/c deflected onto the screen. Higher momentum hadrons or $\gamma$-rays are not deflected onto the screens and are instead absorbed by the beam dump downstream of the electromagnet. An absolute calibration of number of electrons and positrons measured with the magnetic spectrometer is obtained by direct comparison with the amount of light collected from the post-target luminescence screens, taking the different optical setups into account (the shaded regions in Figure~\ref{fig:HRM5+6} account for the errors associated with this absolute calibration). The results show that FLUKA simulations are able to accurately predict both the absolute number and the spectral shape of the e$^{\pm}$ pairs that can reach the screens. In this energy range, the number of pairs is measured to be $N_{\pm,\rm{exp}}=(2.46\pm0.62)\times10^{11}$ and positron fraction is $(N_{\rm{e}^+}/N_{\rm{e}^-})_{\rm{exp}}=0.92\pm0.05$, compared with the FLUKA simulation: $N_{\pm,\rm{FLUKA}}=2.45\times10^{11}$ and $(N_{\rm{e}^+}/N_{\rm{e}^-})_{\rm{FLUKA}}=0.89$. The results of both of the luminescence screen diagnostics are summarized in Table~\ref{tab:numbers_summary}.
\\

Given the experimental validation of the pair beam characteristics predicted by FLUKA simulations, we use the simulations to obtain the beam properties at the rear surface of the target. At this position the pair density is at its highest ($n_{\pm}=\SI{1.64e12}{\per\centi\metre\cubed}$). We assess whether this density is sufficiently high for the pair beam to be considered a plasma. A fundamental characteristic of a plasma is its ability to screen electric potentials that are applied to it. This is the case if the physical size of the plasma exceeds the characteristic Debye screening length, $r_{\rm{scr}}$, or equivalently, if the number of particles per Debye screening volume is much greater than one, i.e. $N_{\pm}/N_{\rm{D}}\gg1$, where $N_{\rm{D}} = n_{\pm}r_{\rm scr}^3$.

The pair beams produced are highly relativistic ($\langle \Gamma \rangle \gg 1$), possessing a relativistic thermal spread ($\Gamma_T =k_{\rm B}T_{\pm}/m_{\rm e}c^2 \gg 1$). Therefore the appropriate screening length is that of a relativistically hot pair plasma, defined by $r_{\rm scr}^{-2} = 8\pi n_{\pm} e^2 / k_{\rm{B}} T_{\pm}$. This is derived assuming the electrons and positrons have a relativistic Maxwellian (Jüttner-Synge) thermal distribution with pair temperature $T_{\pm}$ (see Methods). In the derivation, the plasma has no bulk flow so we evaluate $N_{\pm}/N_{\rm{D}}$ in an inertial frame where the beam has zero net momentum (the `zero-momentum' frame). The pair temperature is obtained in the zero-momentum frame by performing a Lorentz transformation ($\Gamma_0\sim4.5$) on the FLUKA-simulated momentum distributions, leading to an approximately-isotropic momentum distribution which is fit by $T_{\pm}=4~\mbox{MeV}$. The pair density in the zero-momentum frame is given by $n_{\pm}/\Gamma_0$, where $n_{\pm}$ is the pair density in the laboratory frame. Using these parameters, the number of particles in a Debye screening volume is $N_{\pm}/N_{\rm{D}} \sim 8$, with scaling
\begin{equation}
    \frac{N_\pm}{N_{\rm D}}=8~\bigg(\frac{N_{\pm}}{1.5\times10^{13}}\bigg)\bigg(\frac{n_{\pm}}{\SI{1.6e12}{\per\centi\metre\cubed}}\bigg)^{1/2}\bigg(\frac{\Gamma_0}{4.5}\bigg)^{-1/2}\bigg(\frac{T_{\pm}}{\SI{4}{\mega\electronvolt}}\bigg)^{-3/2}.
\end{equation}

The pair beams can be used to study plasma processes relevant to the interpenetration of pair jets within an ambient electron-ion plasma. In this case, the condition to observe collective plasma phenomena is less strict. The beam size must exceed the characteristic growth length of plasma instabilities, which is typically the plasma skin depth of the ambient plasma, $\lambda_{\rm s}^2 = m_{\rm{e}} c^2/4\pi n_{\rm{p}} e^2$. Comparing the pair number with $N_{\rm s} = n_{\pm}\lambda_{\rm s}^3$, assuming equal density for the pair beam and the ambient plasma, we obtain $N_{\pm}/N_{\rm s}\sim132$.

\begin{figure}[t]%
\centering
\includegraphics[width=0.85\textwidth]{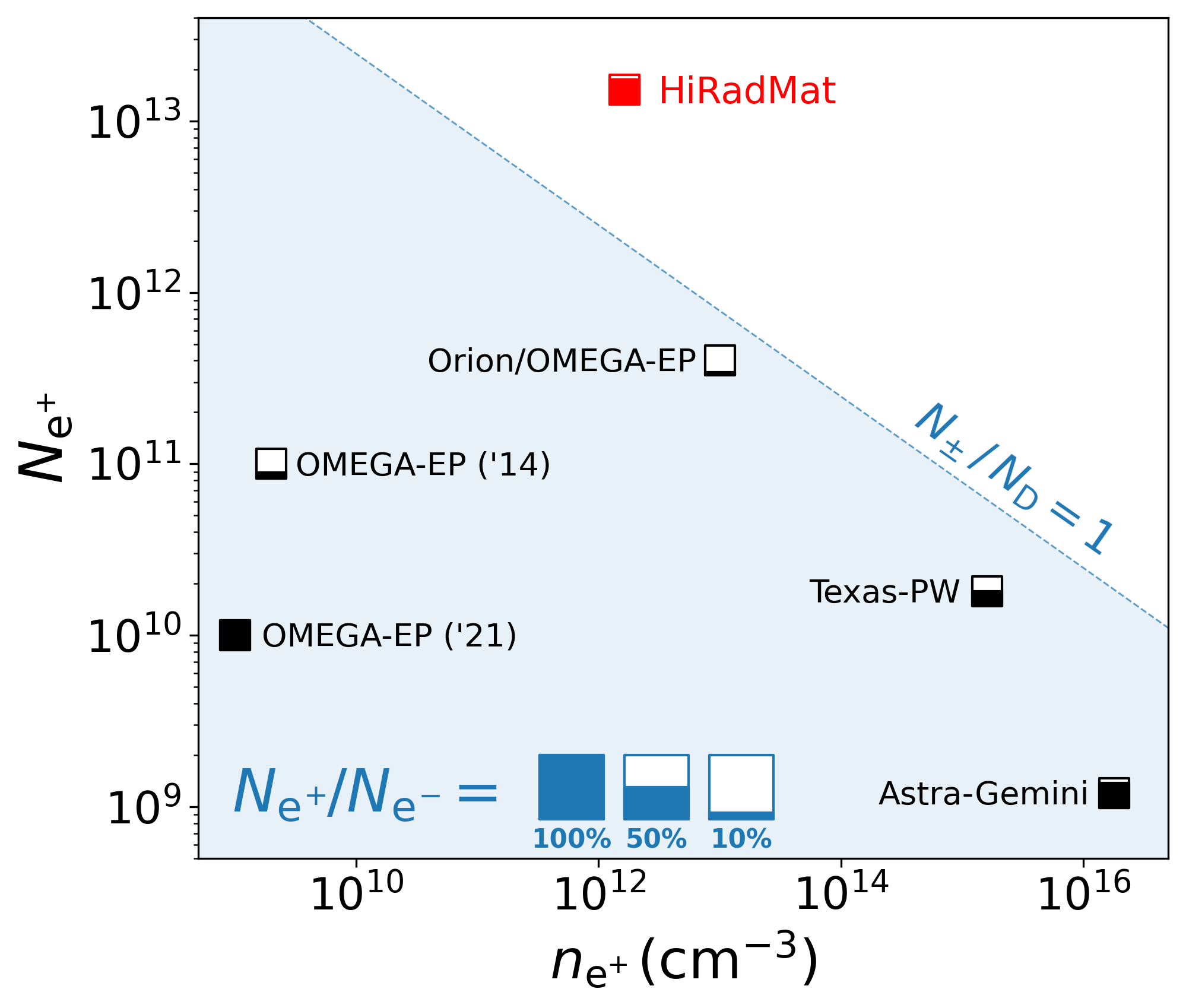}
\caption{{\bf Comparison of laboratory-produced high-density pair beams.} The peak number and density of pairs reported in this study (red square) compared with previous experiments performed at high-power laser facilities (black squares): Orion/OMEGA-EP~\cite{chen2015scaling}, Texas-PW~\cite{liang2015high}, Astra-Gemini~\cite{sarri2015generation}, OMEGA-EP('21)~\cite{peebles2021magnetically}, OMEGA-EP('14)~\cite{chen2014magnetic}. The data are labelled by the facility where the experiment was performed, and the fill fraction of each marker corresponds to the fraction of positrons to electrons in the experiment, $N_{\rm{e}^+}/N_{\rm{e}^-}$ ($\blacksquare=100\%$, $\square=0\%$, also see the key). The blue-shaded region corresponds to when the number of particles per Debye sphere is fewer than one, $N_{\pm}/N_{\rm{D}}<1$ (assuming the screening length used in this work).}
\label{fig:comparison}
\end{figure}

In Figure~\ref{fig:comparison} we plot the number of pairs and beam density achieved in our experiment alongside results of high densities of pairs reported from previous experiments. High pair densities and yields can be achieved in laser experiments, but at the expense of beam dimensions or charge neutrality. 
\\ 

By these definitions, the electron-positron beam we produced at HiRadMat can safely be assumed to behave as a relativistic pair plasma, making it possible to study collective plasma phenomena and providing a laboratory benchmark of several high-energy astrophysics processes.

\begin{table}[t]

{\renewcommand{\arraystretch}{1.7}
\resizebox{1\linewidth}{!}{%
\begin{tabular}{|c|cc|cc|}
\hline
\multirow{2}{*}{Position}                             & \multicolumn{2}{c|}{Pair yield, $N_{\pm}$}        & \multicolumn{2}{c|}{Positron fraction, ($N_{\rm{e}^+}/N_{\rm{e}^-}$)} \\
     & Simulation          & Experiment                          & Simulation            & Experiment                        \\ \hline
Target rear surface ($E \geq 1$~MeV)              & $1.53\times10^{13}$ & - & $0.82$                & -                                 \\
Post-target screen ($E \geq 1$~MeV)              & $1.04\times10^{13}$ & $(1.02\pm0.05)\times10^{13}$ & $0.91$                & -                                 \\
Spectrometer ($30\leq E\mbox{ [MeV]} \leq220$) & $2.45\times10^{11}$ & $(2.46\pm0.62)\times10^{11}$ & $0.89$                & $0.92\pm0.05$            \\ 
\hline
\end{tabular}}}

\caption{{\bf Summary of the measured and simulated yields of electron-positron pairs.} The experimentally measured and FLUKA-simulated electron-positron pair yield ($N_{\pm}$) and positron fraction ($N_{\rm{e}^+}/N_{\rm{e}^-}$) are summarized for the following positions: (i) at the rear surface of the target, (ii) on the post-target luminescence screen, and (iii) on the particle spectrometer luminescence screens.}
\label{tab:numbers_summary}
\end{table}

\clearpage


\section*{Methods}\label{Methods}

\subsection*{Screening lengths of relativistic plasmas}
For a relativistic plasma
the screening length is defined through the static limit of the dielectric tensor $\varepsilon_{ij}$ \cite{Lifshitz1981Physical,silin1960electromagnetic,braaten1993neutrino}:
\begin{equation}
    \lim_{k = 0}\,\lim_{\omega/k = 0} k^2(\varepsilon^l-1) = r_{\rm scr}^{-2},
\end{equation}
where the longitudinal component of the dielectric tensor is
\begin{equation}
    \varepsilon^l = 1 + \sum_{\alpha}\frac{4 \pi q_{\alpha}^2}{k^2} \lim_{\eta\rightarrow+0}\int \mathbf{k} \cdot \frac{\partial f_{\alpha}(\mathbf{p})}{\partial \mathbf{p}} \frac{d^3p}{\omega + i\eta - \mathbf{k}\cdot\mathbf{v}},
\end{equation}
and $\alpha$ labels each species component of the plasma. Assuming a pair plasma comprised of electrons and positrons which have a relativistic Maxwellian (Jüttner-Synge) distribution:
\begin{equation}
    f(\mathbf{p}) = \frac{n_e}{8\pi\Gamma_{T}^3} \exp \left(-\frac{\Gamma}{\Gamma_{T}}\right),
\label{eq:relativisticMaxwellian}
\end{equation}
with ultra-relativistic temperature $\Gamma_T = k_{\rm{B}} T_{\pm}/m_{\rm{e}} c^2 \gg 1$ (noting that this tends to a Maxwellian in the limit $\Gamma \ll 1$, $\Gamma_T \ll 1$). The screening length can be evaluated in the relativistic limit ($v=c$) as
\begin{equation}
    r_{\rm scr}^{-2} = \frac{8\pi n_{\pm} e^2}{k_{\rm{B}} T_{\pm}},
\label{eq:screeninglength}
\end{equation}
where $n_\pm$ and $T_\pm$ refer to the density and temperature of pairs. Identical treatment in the classical limit for an electron-ion plasma leads to the well-known Debye screening, $r_{\rm scr,D}^{-2} = k_{\rm{D}}^2 = 4\pi n_{\rm{e}} e^2/k_{\rm{B}} T_{\rm{e}}$. Since this is derived for a plasma which has net-zero momentum flow, we obtain the plasma density and temperature by fitting Jüttner-Synge distributions in a Lorentz-boosted inertial frame (see the Supplementary Information for details of the fitting).

\subsection*{Electron-positron pair production target}
The target is designed such that a quasi-neutral e$^{\pm}$ beam is produced with pair beam density maximized along a 1-m length downstream of the target. The target can be irradiated by many (potentially hundreds or thousands) of single bunches containing $O(10^{11})$ primary protons, without compromising its structural integrity. In addition, the target must cool sufficiently fast to return to room temperature in-between shots, at a maximum repetition rate of ~1 bunch per minute. The target design has been optimized using simulations with two codes: (i) FLUKA \cite{ahdida2022new,battistoni2015overview} (along with the associated interface ``Flair" \cite{vlachoudis2009flair}), a particle transport Monte-Carlo scattering code capable of accurately describing the hadronic and electromagnetic cascades when the target is irradiated with 440~GeV/c protons, and (ii) Ansys$^{\mbox{\textregistered}}$ Mechanical \cite{ansysmanual}, a finite-element code used to estimate the target cooling and the stress/strain induced by the energy deposition of the beam. A FLUKA-simulated transient thermal field is used as the initial condition. 

FLUKA uses a robustly bench-marked physics model. To achieve a good statistical representation in the Monte Carlo method, $10^5$ iterations were performed. The low-energy cutoff for particle transport in the simulation is $10$~keV for $\rm{e}^-/\rm{e}^+/\gamma$ and $100$~keV for hadrons.

The chosen target design consisted of a 360~mm long cylinder of isostatic graphite (SGL Carbon R6650, \SI{1.84}{\gram\per\cubic\centi\metre}) and 10-mm thick disk of tantalum, both having a $\SI{20}{\milli\metre}$ diameter. The graphite and tantalum are housed inside a 400-mm length, 50-mm diameter cylinder of high-strength T9 aluminium alloy that acts as both a confinement vessel and a heat sink. The tantalum is press-fit to ensure maximal thermal contact. 2-mm thick expanded graphite pieces (SGL Carbon Sigraflex, \SI{1}{\gram\per\cubic\centi\metre}) separate the target components to allow thermal expansion and reduce contact stresses during irradiation, while 2-mm thickness Sigradur G glassy carbon beam windows are clamped onto either end of the target by aluminium flanges with Viton O-rings to hermetically seal the target materials. Using this design, the Ansys simulations have shown that the thermal loading per $3\times 10^{11}$ protons is highest inside the tantalum, reaching peak instantaneous temperatures of $\SI{300}{\celsius}$. Radiative and convective cooling via the outer surface of the target housing leads to cooling of the target to room temperature within a few seconds following the beam impact, while the beam-induced maximum strain of the tantalum remains in all cases well below its plastic deformation limit.

\subsection*{Chromium-doped luminescence screens}

Chromium-doped alumina-ceramic luminescence screens (Chromox, Al$_2$O$_3$:~99.5\%, Cr$_2$O$_3$:~0.5\%~\cite{mccarthy2002characterization}) have been used to measure the particle beam intensity and transverse profile during the experiment. In the Chromox screens, principal luminescence is due to de-excitations of the lowest-excited state of Cr$^{3+}$ when energy is deposited in the screen by ionizing particles and radiation. Light is emitted isotropically, strongest at wavelengths $\lambda_1 = 691$~nm, and $\lambda_2=694$~nm with decay times $3-6$~ms. 

The transverse beam profile of the secondary beam is imaged using a $70$~mm~$\times\, 50$~mm~$\times \, 0.25$~mm screen positioned 10~cm downstream of the target, and a blocker foil (\SI{50}{\micro\metre} aluminium) is used to minimize stray optical light. The screen is oriented at 45$^{\circ}$ to the beam path and viewed directly by a digital camera (Basler acA1920-40gm GigE camera with Sony IMX249 CMOS sensor and Canon EF 75-300 mm f/4-5.6 III lens) at a standoff distance 3.8~m with an exposure time 24~ms. An almost identical optical setup is used to image the screens in the magnetic spectrometer, except larger screens are used ($200$~mm~$\times\, 50$~mm~$\times \, 1$~mm, centred at a distance $240$~mm off-axis), and viewed through a single mirror reflection at a standoff distance 6.2~m. 

Given that relativistic particles in the energy range of interest exhibit minimum-ionizing behaviour, the energy deposition of a particle passing through the Chromox screen is expected to be approximately insensitive to energy and constant between singly-charged particle species (a result confirmed by FLUKA simulations, see the Supplementary Information).
The translucence of the Chromox screens to the luminescence light (attenuation length, $\mu=0.8$~mm$^{-1}$) limits the spatial resolution to $\gtrsim$\SI{100}{\micro\metre}, as the luminescence light is not significantly attenuated as it is transmitted from a region where energy is deposited deeper into the screen. The translucence of the screens simplifies the analysis; as a first approximation we don't consider the different longitudinal energy deposition profiles, which are anyway shown in simulations to be approximately uniform through the screen thickness for the relativistic particles observed.

\subsection*{Magnetic electron-positron spectrometer}

Before the experiment, we have characterized the spatial magnetic field profile of the electromagnet with currents supplied to the coils in the range between $0-400$~A. This electromagnet, designated ``MNPA" in the CERN internal naming system, has a yoke length of $250$~mm (total length $544$~mm), an aperture of width $260$~mm and height $202$~mm, and a maximum field at 400A of 0.34 T. The exact magnet geometry has been modelled in detail using the finite-element code Opera 3D \cite{opera3d}. The field map has been calculated with sufficient resolution (10~mm) to capture the magnetic field gradients inside the magnet gap, given that magnetic field transitions from 10\%-90\% peak field strength over a distance $200$~mm. The field map has been cross-checked against direct measurements of the magnet field using a Hall probe and the difference between the model and measured magnetic fields inside the magnet gap are $\lesssim2$\%. Finally, an energy calibration for the spectrometer (correlating screen position with e$^{\pm}$ energy) is obtained from particle ray-tracing calculations using the magnetic field maps. The e$^{\pm}$ energy ranges sampled by different magnet settings are $30-55$~MeV, $55-80$~MeV, $75-110$~MeV, and $90-220$~MeV.

An absolute calibration of the electron and positron numbers is made by using the brightness (pixel counts per unit area) of the in-beam luminescence screen. 
Specifically, we account for the difference between the amount of light collected in the optical setups used for the in-beam screen luminescence and for the spectrometer screens, considering (i) the different standoff distances, which leads to different solid angle subtended ($d\Omega_{\rm beam}/d\Omega_{\rm spect}=2.7\pm0.4$); (ii) the different camera gain settings used ($G_{\rm spect} = 38 \pm 2$, $G_{\rm beam} = 1$); and (iii) the different thickness of screen used, where thicker screens lead to larger energy deposition per particle ($u_{\rm dep,spect}/u_{\rm dep,beam}=3.5\pm0.5$). 
A small $1/\cos \theta$ geometric correction is applied to the energy spectra to account for the additional path length of Chromox encountered by obliquely-incident deflected particles, where $\theta=10^{\circ}$ at the screen edge closest to the beam axis, and $\theta=25^{\circ}$ at the furthest edge.

The digital cameras viewing the in-beam screen at a standoff distance of $\SI{3.8}{\metre}$ can resolve features as small as $\SI{50}{\micro\metre}$ in size, whilst the cameras viewing the spectrometer screens at a standoff distance $\SI{6.2}{\metre}$ can resolve features $\SI{120}{\micro\metre}$ in size. However, the resolution of the energy spectrum projected onto the spectrometer screens is limited by the 20-cm thickness, 20-mm wide concrete aperture at the entrance of the electromagnet.

In our experimental setup, the target has been placed on a vertically movable high-precision stage, allowing us to acquire data with the target in-beam, as well as in a `target-out' position. In the latter case, the primary proton beam continues at its full intensity through the luminescence screens and the  electromagnet towards the beam dump. Using the `target-out' configuration, we took measurements without the current supplied to the electromagnet to characterize the hadron and lepton background produced as particles are back-scattered by the proton beam impact on the beam dump. The remnant field of the aforementioned electromagnet was measured extensively before the experiment using a Hall-probe and was found reproducibly to be negligible (on the order of the noise of the instrument, i.e. $B\lesssim\SI{0.3}{\milli\tesla}$). 




\backmatter





\bmhead{Acknowledgments}
We thank Prof.~C.~Joshi~(UCLA), Dr.~F.~Albert~(LLNL) and Dr.~C.~Densham (STFC Rutherford Appleton Laboratory) for useful discussions, and Dr.~T.~Ma~(LLNL) for supporting this experiment. This project has received funding from the European Union’s Horizon Europe Research and Innovation programme under Grant Agreement No 101057511 (EURO-LABS).
The work of G.G. was partially supported by UKRI under grants no. ST/W000903/1 and EP/Y035038/1, while A.F.A.B. was also supported by UKRI (grant number MR/W006723/1). The work of D.H.F. and D.H. was supported by the U.S. Department of Energy under Award Number DE-NA0004144.
We also acknowledge funding from AWE plc., and the Central Laser Facility (STFC). 
FLUKA simulations were performed using the STFC Scientific Computing Department's SCARF cluster. UK Ministry of Defence © Crown Owned Copyright 2023/AWE.

\bmhead{Author contributions}
This project was conceived by G.G. and R.B. The experiment was designed by C.D.A., G.G., P.S., N.C. and R.B., and carried out by C.D.A., P.S., N.C., G.G., T.H., R.S., J.H., P.B., S.B., F.D.C., A.M.G., D.H., S.I., V.S., T.V. and B.T.H. The data analysis was carried out by C.D.A. The manuscript was written by C.D.A., with input from G.G., N.C. and R.B. Numerical simulations were performed by C.D.A. and P.S. Further experimental and theoretical support was provided by I.E., D.H.F., A.F.A.B., A.A.S., J.T.G., B.T.H., F.M., S.S., B.R., H.C., L.O.S., T.D. and R.M.G.M.T. 










\newpage
\section*{Supplementary information}
\subsection*{Detailed secondary beam characteristics from FLUKA simulations}
\begin{figure}[h!]%
\centering
\includegraphics[width=0.55\textwidth]{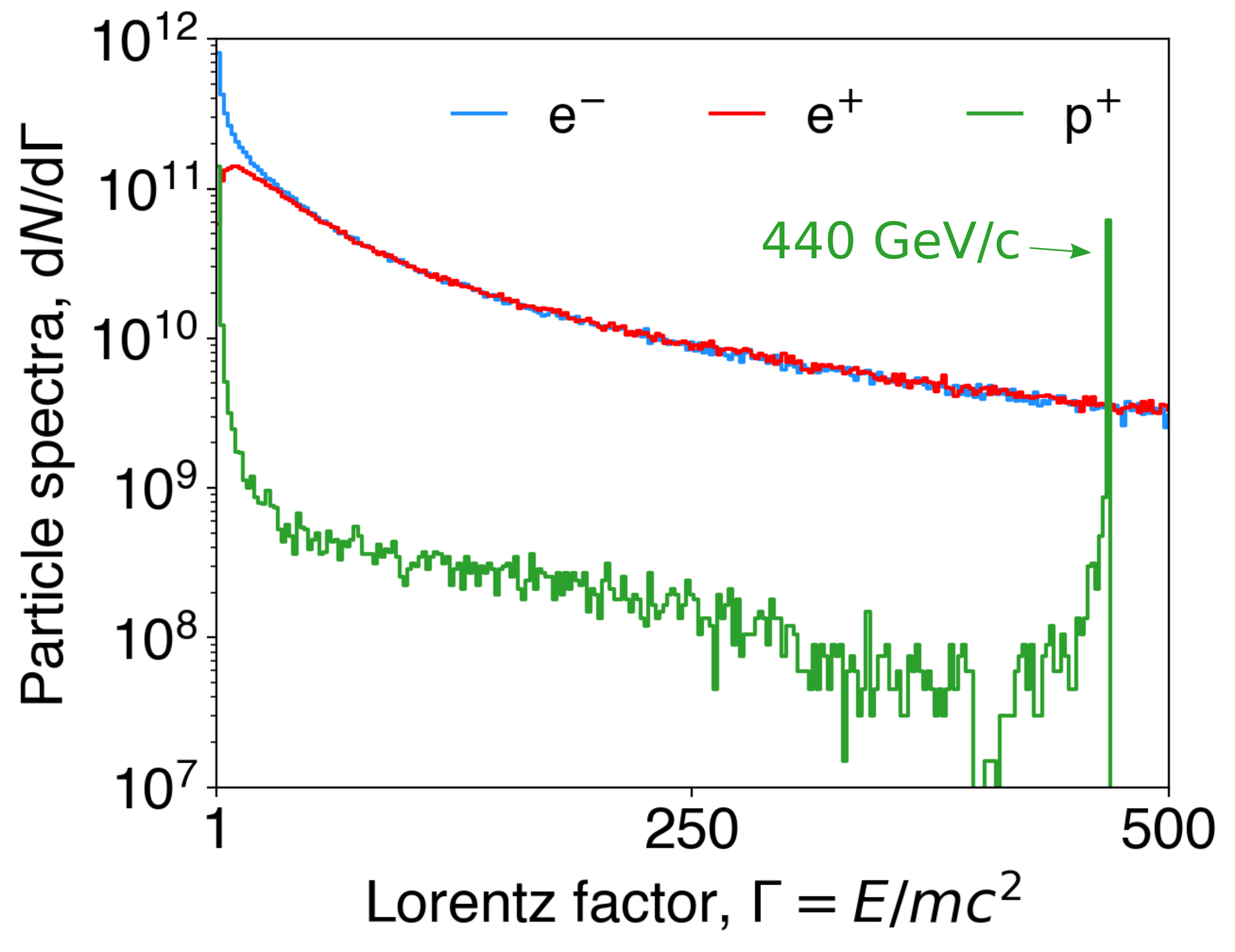}
\caption{{\bf FLUKA simulations of particle spectra.} Particle spectra of electrons (e$^{-}$, blue), positrons (e$^{+}$, red) and protons (p$^{+}$, green) at the exit of the target are obtained from FLUKA simulations and plotted as a function of relativistic Lorentz factor, $\Gamma = E/mc^2$. The electron and positron spectra are matched except at energies $\lesssim\SI{10}{\mega\electronvolt}$, where electrons can more easily escape the target than positrons. The e$^{-}$e$^{+}$ spectra are characterized by extended tail distribution functions with energies up to several tens of GeV. The population of primary protons which experience only elastic scattering are identified by the peak in the spectra corresponding to proton momentum $p=\SI{440}{\giga\electronvolt}$/c.}
\label{fig:FLUKA_particle_spectra_edit}
\end{figure}

The yield, mean energy and beam radius of all secondary species produced by the proton beam irradiation of the target are given in Table~\ref{tab:FLUKA_particle_parameters}, obtained at the rear surface of the target. The particle spectra of electrons, positrons and protons are given as a function of relativistic Lorentz factor in Figure~\ref{fig:FLUKA_particle_spectra_edit}.

\begin{table}[h]

{\renewcommand{\arraystretch}{1.25}
\resizebox{\linewidth}{!}{%

\begin{tabular}{|c|c|c|c|c|c|}
\hline
Species & Specified energy range & Yield (per primary)  & Mean energy, $\langle E \rangle$ (GeV) & Beam radius, $\sigma_r$ (mm) \\   \hline
e$^-$                                      & $\geq\SI{1}{\mega\electronvolt}$ & 56                & 0.20      & 2.0            \\
e$^+$                                      & $\geq\SI{1}{\mega\electronvolt}$ & 46                & 0.27      & 1.9            \\
p$^+$                                      & $\geq\SI{100}{\kilo\electronvolt}$, $<\SI{430}{\giga\electronvolt}$  & 1.4    &    31     &   1.2         \\
p$^+$                                    &  $\geq\SI{430}{\giga\electronvolt}$ & 0.42            & 439     &   1.2         \\

$\pi^{\pm}$                             & $\geq\SI{100}{\kilo\electronvolt}$  & 6.2      &    6.6 & 1.7                 \\
k$^{\pm}$                                &  $\geq\SI{100}{\kilo\electronvolt}$  & 0.5                 & 11      &    1.4        \\
µ$^-$                                   & $\geq\SI{100}{\kilo\electronvolt}$  & $<$0.05                & -    &    -          \\
$\gamma$                                 & $\geq\SI{10}{\kilo\electronvolt}$   & 600                 & 0.07 & 2.0                  \\
n                                     &  $\geq\SI{100}{\kilo\electronvolt}$   & 12.6                & 3.1     &  2.2           \\ \hline
\end{tabular}}}
\caption{{\bf Detailed secondary beam characteristics from FLUKA simulations.} The yield (per primary), mean energy ($\langle E \rangle = \langle \Gamma \rangle mc^2$), and beam radius (half-width-half-maximum) are provided for the most abundantly produced secondary species: electrons (e$^{-}$), positrons (e$^{+}$), protons (p$^{+}$), pions ($\pi^{\pm}$), kaons (k$^{\pm}$), muons (µ$^{-}$), $\gamma$-rays and neutrons (n). The specified energy range defines the range of energies that the table parameters refer to. In the case of electrons and positrons, for ease of comparison with other studies, the low-energy cutoff used is $\SI{1}{\mega\electronvolt}$, though there are even more pairs (up to 10\% more) if the energy cutoff is reduced to \SI{10}{\kilo\electronvolt}. For the remaining species, the low energy cutoff is defined by the simulation. The yield per primary can be multiplied by the primary proton beam intensity to obtain the actual yield. For instance, with a primary proton beam intensity of $3\times10^{11}$ protons, a total of $1.7\times10^{13}$ electrons and $1.4\times10^{13}$ positrons are produced. For the muons, the simulated number is too small for reasonable statistics of the mean energy and beam radius. }
\label{tab:FLUKA_particle_parameters}
\end{table}

\newpage
\subsection*{Chromox luminescence as a function of particle type and energy}

\begin{figure}[h]%
\centering
\includegraphics[width=0.7\textwidth]{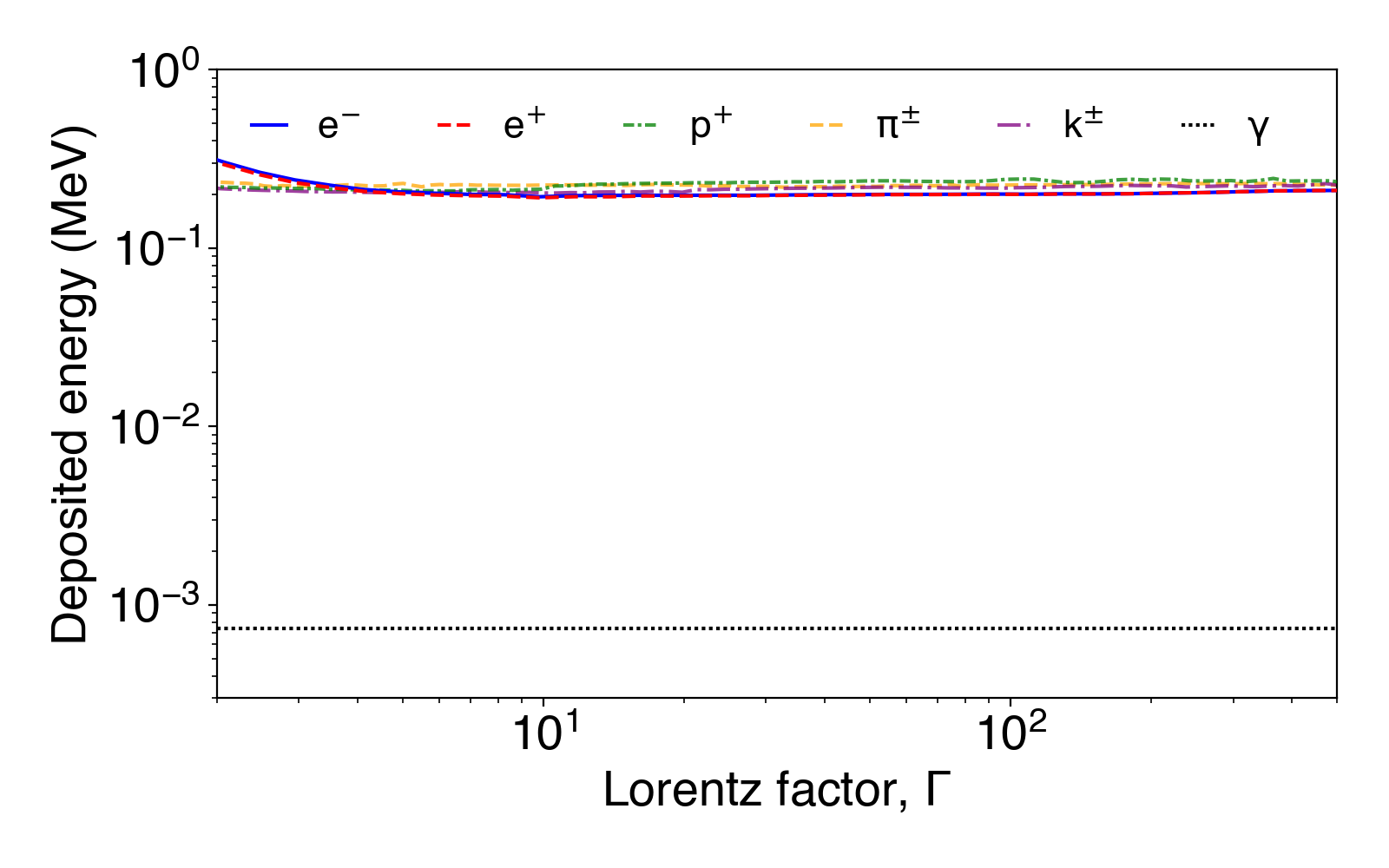}
\caption{{\bf FLUKA simulations of energy deposited into Chromox screens.} The energy deposited per particle into a \SI{250}{\micro\metre}-thickness Chromox screen at \SI{45}{\degree} incidence is obtained from FLUKA simulations. Deposited energy is plotted as a function of relativistic Lorentz factor for: electrons (e$^{-}$, blue-solid), positrons (e$^{+}$, red-dashed), protons (p$^{+}$, green-dot-dashed), charged pions ($\mathrm{\pi}^{\pm}$, orange-dashed) and charged kaons (k$^{\pm}$, purple-dot-dashed). The mean energy deposited per $\gamma$-ray in the range MeV-TeV is also plotted (black-dotted), showing that the energy deposition of $\gamma$-rays is approximately $300\times$ smaller.}
\label{fig:chromox250}
\end{figure}

To further validate the assumption that particles deposit an approximately equal amount of energy into the screens, FLUKA simulations are performed, irradiating Chromox with different particle types and energies. The results are shown in Figure~\ref{fig:chromox250}. In the range $2\lesssim\Gamma\lesssim500$, the deposited energy per particle is approximately the same across energy scales and between charged particle species. The mean energy deposited for $\gamma$-rays in the range MeV-TeV is plotted (black-dotted), showing that the energy deposition per $\gamma$-ray is approximately $300\times$ smaller. Therefore, we expect the contribution of $\gamma$-rays to observed screen luminescence to be negligible, despite the expected yield being $6\times$ greater than e$^{+}$e$^{-}$.

\subsection*{Magnetic particle spectrometer raw data}
\begin{figure}[h]
\centering
\includegraphics[width=0.6\textwidth]{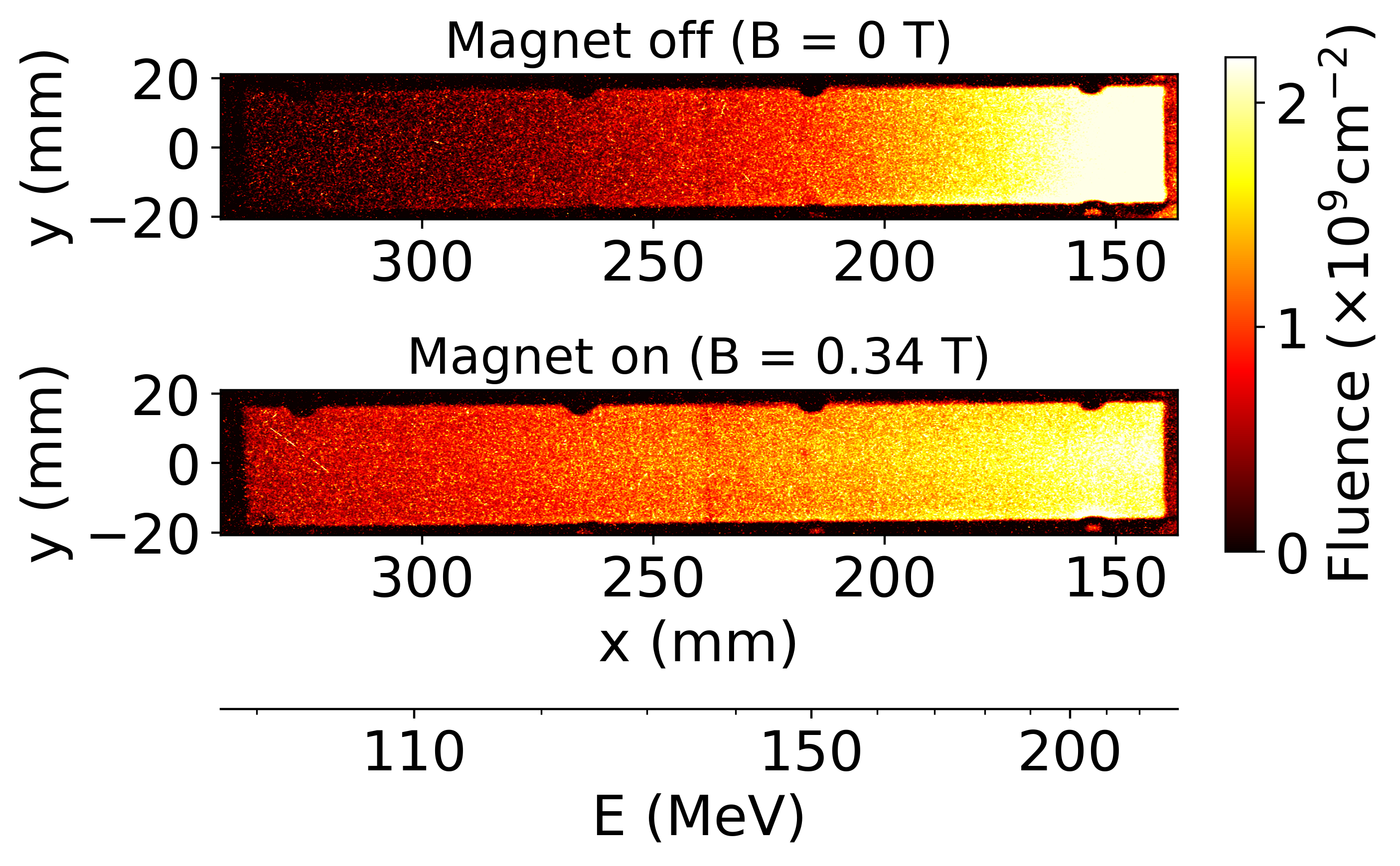}
\caption{{\bf Magnetic particle spectrometer raw data.} Raw image data of the chromium-doped luminescence screens used to measure the energy spectra of electrons and positrons in the magnetic particle spectrometer. The screen corresponding to positrons is shown when the target is irradiated and the electromagnet is turned off (upper), and when the electromagnet is turned on with peak magnetic field $B=\SI{0.34}{\tesla}$ (lower). When the magnet is turned on, positrons are deflected away from the beam axis onto the screen. The secondary x-axis shows the corresponding positron energy according to the energy calibration when the magnet is operating at the given setting.}
\label{fig:HRM5_raw}
\end{figure}

Raw image data of the Chromox screens in the magnetic particle spectrometer is shown in Figure~\ref{fig:HRM5_raw} (post-background subtraction). The increase in signal when the electromagnet is turned on is evidence of positrons in the beam with energy $E \leq \SI{220}{\mega\electronvolt}$. The horizontal lineouts from this data in the central 25 mm are used to obtain the electron and positron spectra shown in Figure~\ref{fig:HRM5+6}.

\newpage

\subsection*{Fitting a relativistic Maxwellian to the pair beam momentum distribution}

\begin{figure}[h!]%
\centering
\includegraphics[width=1\textwidth]{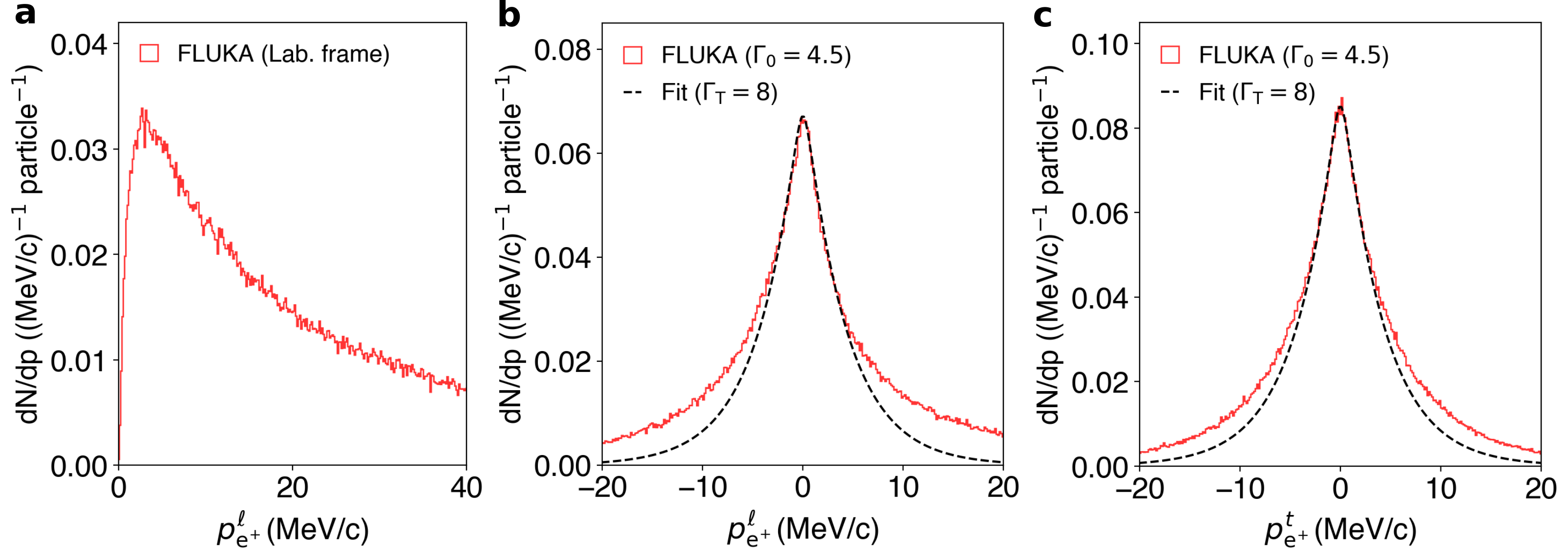}
\caption{{\bf Fitting the momenta distributions to relativistic Maxwellian distributions.} (a) The longitudinal momentum distribution of positrons, $p_{\rm{e}^+}^{\ell}$, corresponding to the laboratory frame ($\Gamma_0=1$) is obtained from FLUKA simulations (red). (b) Once a Lorentz transformation ($\Gamma_0 = 4.5$) is applied to the particle vectors, the longitudinal momentum spectrum is modified such that it becomes approximately symmetric about zero-momentum. A relativistic Maxwellian distribution is fitted with temperature parameter $\Gamma_T=8$ (black-dashed). (c) The transverse momentum distribution of positrons, $p_{\rm{e}^+}^{t}$, is plotted with the same fit.}
\label{fig:fittingmomenta}
\end{figure}

Given that the pair plasma beams produced in this work are characterized by non-Maxwellian distribution functions, possessing relativistic thermal spreads and bulk flow ($k_{\rm B}T_{\pm} \gg m_{\rm e} c^2$), comparisons with lengthscales traditionally used to define collective behaviour in classical plasmas are done with care. The pair temperature ($T_{\pm}$) is identified as a key parameter which must be included in the consideration of the limiting scales of collective plasma processes. 
\\
\\
We derive the screening length for a relativistically hot plasma assuming a relativistic Maxwellian (Jüttner-Synge) distribution (Eq.~\ref{eq:relativisticMaxwellian}). This screening lengthscale is analogous to the familiar Debye screening length for non-relativistic Maxwellian plasmas. Since the screening length is derived assuming zero bulk motion, to obtain the appropriate pair temperature we perform a Lorentz transformation on the FLUKA-simulated momentum spectra and fit the data to a Jüttner-Synge distribution. The number of pairs per Debye sphere is calculated in the zero-momentum frame, using the beam density and dimensions consistent with the Lorentz transformation from the laboratory frame. The fitting of the Lorentz-transformed momentum distributions is shown in Figure~\ref{fig:fittingmomenta}. 
\\
\\
The number of pairs per Debye sphere is found to scale with:
\begin{equation}
    \frac{N_{\pm}}{N_{\rm D}} \propto N_{\pm}~{n_{\pm}}^{1/2}~\Gamma_0^{-1/2}~\Gamma_T^{-3/2},
\end{equation}
where $\Gamma_0$ is the Lorentz factor required for a transformation from the laboratory frame to the zero-momentum frame, and $\Gamma_T$ is the Lorentz factor associated with the pair temperature fitted in the zero-momentum frame ($k_{\rm{B}}T_{\pm} = \Gamma_T m_{\rm e} c^2$).
\\
\\
Alternative approaches of comparing the pair beam with limiting scales of plasma collective processes do not consider the thermal spread of the beam in the screening length. Instead, a comparison is made with the plasma skin depth, using a density corresponding to a Lorentz boost to a frame co-moving with the mean pair energy ($\langle E\rangle$ = $\langle\Gamma\rangle m_{\rm e} c^2$) \cite{chen2023perspectives}. There are two issues with this approach. The first is that the pair temperature is not considered, and so the treatment is as if the momentum distribution is mono-energetic at the mean particle energy. Secondly, particle spectra with extended tails can have mean Lorentz factor far-in-excess of the median Lorentz factor, therefore misrepresenting the bulk. This alternative approach leads to a different scaling for the number of particles per screening volume: $(N_{\pm}/N_{\rm D})\propto N_{\pm}~{n_{\pm}}^{1/2}~\langle\Gamma\rangle^{-3/2}$. However, the approach presented in this work leads to approximately the same result if $\Gamma_0, \Gamma_T \sim \langle\Gamma\rangle^{3/4}$. Indeed this the case given $\Gamma_0 = 4.5$ and $\Gamma_T = 8$, and provided $\langle \Gamma \rangle \sim 10-20$, which is generally true for the laser experiments compared in Figure~\ref{fig:comparison}.


\end{document}